\documentclass[]{article}
\usepackage{epsfig}
\usepackage{latexsym}

\makeatletter
\def\section{\@startsection{section}{1}{\z@}{-3.5ex plus-1ex minus-.5ex}%
{1.5ex plus.3ex}{\bfseries\large}}
\makeatother
\begin{document}
\begin{center}
{\LARGE Quantum-Hall to insulator transition}\\[.75\baselineskip]
{\large H. Potempa, A. B\"aker, and L. Schweitzer}\\[.5\baselineskip]
{\large\itshape Physikalisch-Technische Bundesanstalt, Bundesallee 100,\\
D-38116 Braunschweig, Germany}
\end{center}

\begin{abstract}
The crossover from the quantum Hall regime to the Hall-insulator is 
investigated by varying the strength of the diagonal disorder in a 2d 
tight-binding model. The Hall and longitudinal conductivities and the 
behavior of the critical states are calculated numerically. 
We find that  with increasing disorder the current carrying states close to 
the band center disappear first. Simultaneously, the quantized Hall 
conductivity drops monotonically to zero also from higher quantized values.
\end{abstract}

\section{Introduction}
The disappearance of the integer quantum Hall effect (QHE) at low magnetic
fields or, equivalently, for strong disorder potentials, has been 
the subject of continuing interest since its discovery.
Based on the continuum model where the energy spectrum is bounded only from 
below it was argued \cite{Khm84,Lau84} that the current carrying states  
continuously float up in energy when the magnetic field is decreased. 
The system becomes an insulator when the last critical state crosses the 
Fermi level. Therefore, a direct transition from the QHE-state to an 
insulating state is possible only from the lowest Landau band \cite{KLZ92}.

On the other hand, for a disordered tight-binding model (TBM) with lattice 
constant $a$ this scenario seems not to be adequate. Here a symmetric 
energy band is formed which splits into $q$ sub-bands when a perpendicular 
magnetic field $B$ is applied which can be expressed by the number of flux 
quanta per lattice cell, $\alpha=eBa^2/h=1/q$, \cite{SKM84}. 
As in the Landau bands of the continuum model the localization length diverges
near the center of the disorder broadened sub-bands \cite{AA81,SKM84,HK90}.
These critical states exhibit a topological property, a nonzero Chern integer
that leads to a quantized Hall conductivity \cite{TKNN82}.  

Recently, it was found for the TBM \cite{LXN96} that with increasing disorder 
the critical states disappear one after another without changing their 
position in energy, starting with the states close to the band center. 
The reason for this behavior is that states with negative Chern number  
originally situated only near the band center move down in energy and 
annihilate those with Chern number $+1$ located in each sub-band.
As a consequence of this scenario a direct QHE to insulator transition from
higher Landau bands becomes possible \cite{SW97,SW98} which has 
already been observed in experiments \cite{JJWH93,Wea94,KMFP95,Sea97a}.

However, from a similar numerical investigation it was concluded \cite{YB96}  
that for weak fields the crossover to the insulator is due to a floating up
of critical levels into the band center, because a magnetic field 
independent critical disorder $W_c\approx 6\,V$ was found where all current 
carrying states disappear. 

In contrast to this view Sheng and Weng \cite{SW97} inferred from their 
calculation of the thermodynamic localization length that the critical disorder
$W_c/V\sim \sqrt{\alpha}$ tends to zero as the magnetic field is lowered. 
This implies that for fixed magnetic field the longitudinal conductivity 
$\sigma_{xx}$ increases in the tails of the sub-bands with increasing 
disorder strength. This increase of $\sigma_{xx}$ should 
also go along with a decrease of the width of the quantum Hall plateaus.  

In this paper, we investigate numerically this scenario by calculating the 
Hall and longitudinal conductivity using a recursive Greens function 
technique \cite{SKM85,Mac85} and extract the behavior of the current 
carrying states from the energy level statistics \cite{OO95,BS97}. 

\begin{figure}
\centering\epsfxsize11.5cm\epsfbox{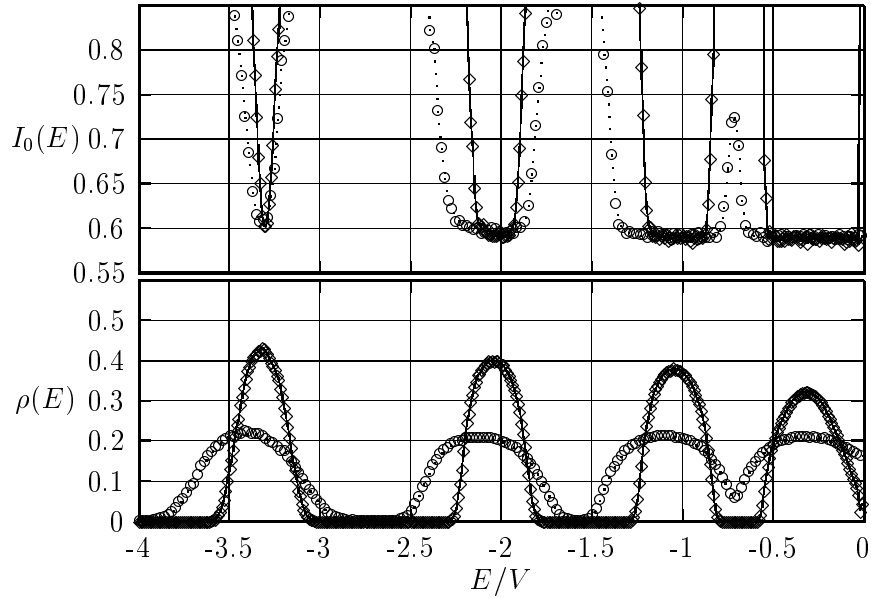}
\caption[]{The energy dependence of the second moment 
$I_0=1/2\,\langle s^2\rangle$ (upper part) and the density of states 
$\rho(E)$ (lower part) for disorder $W/V=1$ ($\diamond$) and $W/V=2$ 
($\circ$).}
\end{figure}

\section{Model and Method}
The lattice model describing non-interacting electrons moving in a 
two-dimensional disorder potential and a perpendicular magnetic field
is defined \cite{SKM84} by an Anderson Hamiltonian where the magnetic 
field is incorporated into the transfer terms $V_{mn}$ via Peierls phase 
factors,
\begin{equation}
H=\sum_{m}\varepsilon_m |m\rangle\langle m| + 
\sum_{<m \ne n>} V_{mn} |m\rangle\langle n|.
\end{equation}
The $|m\rangle$ are lattice vectors associated with the sites $m$ of a 
simple square lattice with spacing $a$. The transfer is restricted to 
nearest neighbors only,  
$V_{mn}=V\exp(i2\pi (e/h) \int_{\mathbf{r}_m}^{\mathbf{r}_n} 
\mathbf{A}(\mathbf{r})\,\mathrm{d}\mathbf{r})$, and for the vector potential 
the Landau gauge, $\mathbf{A}=(0,-B\mathbf{x},0)$, is chosen.  
The disorder potentials  
$\{\varepsilon_m\}$ are represented by a set of independent random numbers
with uniform distribution, $|\varepsilon_m|\le W/2$, where $W$ is the disorder
strength.  

The density of states, the Hall and longitudinal conductivities
are calculated using a recursive Greens function technique developed 
previously \cite{SKM85,Mac85}. This method allows for a computation of large 
lattice sizes. In the present work we take $\alpha=1/8$ and consider sample 
widths $M$ up to $96\,a$ and lengths in the range $10^4\,a \le L \le 10^5\,a$. 
For $\rho(E)$ and $\sigma_{xx}$ periodic boundary conditions are applied
across the sample while for $\sigma_{xy}$ Dirichlet boundary conditions 
are chosen.

The eigenvalues $\{E_i\}$ used for the energy spacing distribution $P(s)$ 
have been calculated by direct diagonalization of square systems using a 
Lanczos algorithm. Here, periodic boundary conditions are applied in both 
directions and the maximal size considered was $M\times M=(128\,a)^2$. 
A large number of realizations were computed resulting in a total of 
more than $5\cdot 10^5$ eigenvalues per disorder strength.
As usual the spacing of consecutive levels is divided by the mean level 
spacing, $|E_i-E_{i+1}|/\Delta\equiv s$, and the spectrum is properly 
unfolded. 
The second moment of the nearest neighbor level spacing distribution, 
$\langle s^2\rangle=\int_0^\infty s^2 P(s)\,\mathrm{d}s$, serves as a 
measure of the electron localization length. We calculate the quantity
$I_0=1/2\,\langle s^2\rangle$ which in the thermodynamic limit is known to be
$I_0^{loc}=1$ for strongly localized states, and from random matrix theory 
$I_0^{ext}=0.59$ for a metallic system with unitary symmetry. 
For the QHE system only localized and critical states are to be expected. 
The latter are easily distinguished because the level statistics is scale 
independent at the critical point \cite{Sea93,OO95,BS97}.

\section{Results and Discussion}
In Fig.~1 $I_0=1/2\,\langle s^2\rangle$ is shown (upper part) as a function of 
energy for disorder strength $W/V=1$ and 2.  
The corresponding density of states (DOS) is plotted below. Because of the 
symmetry with respect to $E/V=0$ only half of the energy range is displayed.
Increasing the disorder broadens the DOS and also the energy range where the 
localization length of the electronic states exceeds the system size.
The 3rd and 4th minima of $I_0$ start to merge already at disorder $W/V=2$ 
which eventually results in a disappearance of these critical states while 
the two lowest minima remain separated up to $W/V=5$.

\begin{figure}
\centering\epsfxsize11.5cm\epsfbox{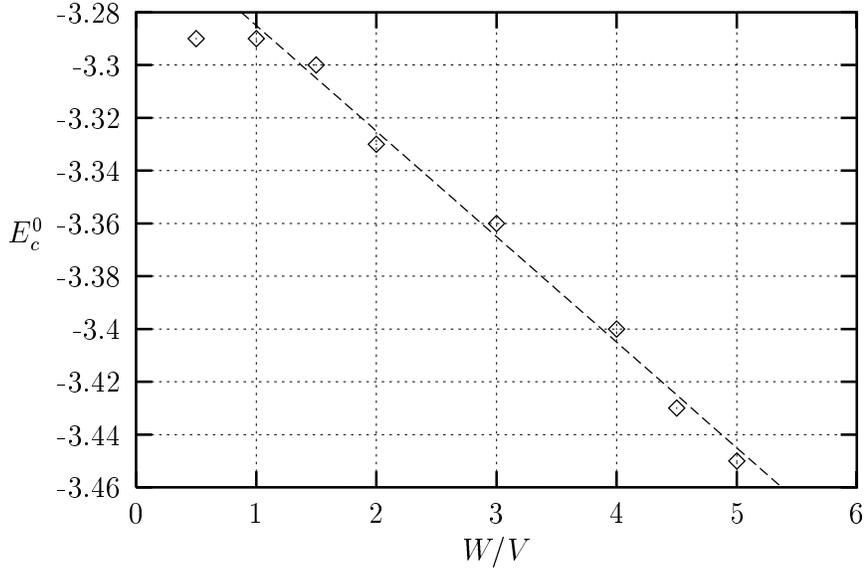}
\caption[]{The disorder dependence of the critical energy $E_c^0$ 
in the lowest Landau band.}
\end{figure}

For the lowest Landau band, the disorder dependence of the position 
of the minimum of $I_0(E)$, $E_c^0$, is shown in Fig.~2 which 
coincides with the divergence of the localization length. 
A linear shift of $E_c^0$ down to smaller energies is seen 
which corresponds to the broadening of the total tight binding band. 

An opposed shift within the lowest Landau band can be observed if one looks 
at the position of the critical filling factor $\nu_c^0$ instead, which also 
seems to be a more physical quantity than the critical energy when comparing 
with experiment. 
For $W/V>2$, a linear increase of $\nu_c^0$ can clearly be seen in Fig.~3. 
This shift of the critical states to higher filling factors is understood 
as an effect of overlapping Landau bands \cite{LXN96,YB96,SW97}.
The flattening for smaller disorders is due to a peculiar asymmetry of 
the Landau bands in the lattice model.

\begin{figure}[b]
\centering\epsfxsize11.5cm\epsfbox{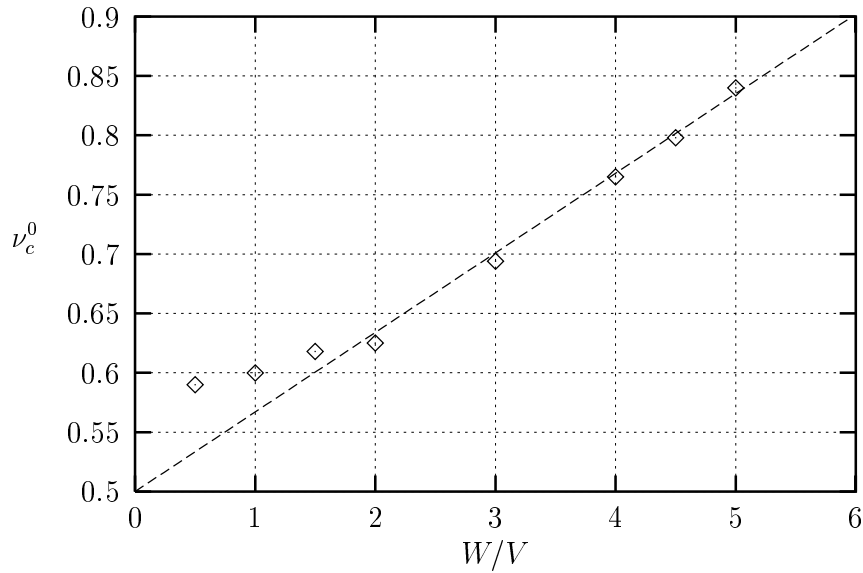}
\caption[]{Disorder dependence of the critical filling factor 
$\nu_c^0$ in the lowest Landau band.}
\end{figure}

\begin{figure}
\centering\epsfxsize11.5cm\epsfbox{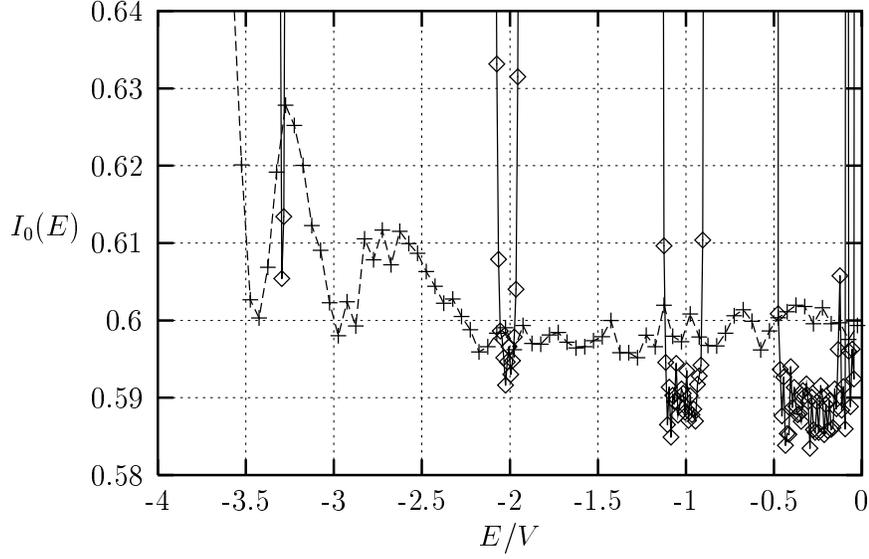}
\caption[]{The energy dependence of the second moment 
$I_0=1/2\,\langle s^2\rangle$ 
for disorder strengths $W/V=0.5$ ($\diamond$) and $W/V=5$ ($+$).}
\end{figure}

However, the opposite changes of $E_c^0$ and $\nu_c^0$ finally prove to be 
irrelevant 
for the fate of the current carrying states because the shifts are small 
compared to the downward movement of critical states from the band center 
that cause the disappearance of the current carrying states of the higher 
Landau bands. 
The imminent destruction of the last current carrying state is
illustrated in Fig.~4 where $I_0(E)$ is plotted for disorder strengths 
$W/V=0.5$ and 5.
All minima of $I_0(E)$ exhibit scale invariant behavior for $W/V=0.5$ as 
expected for critical states. For $W/V=5.0$, however,
only the two lowest minima remain scale invariant while the states are
scale dependent for energies larger than $\approx -2.8$. Increasing the 
system size shifts the value of $I_0$ to 1 as expected for localized states.

\begin{figure}
\centering\epsfxsize11.5cm\epsfbox{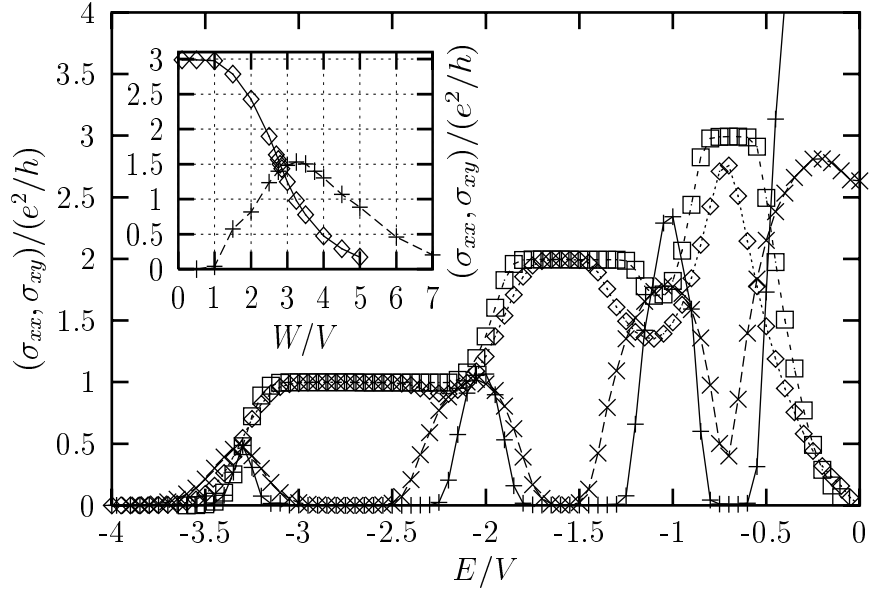}
\caption[]{The Hall conductivity $\sigma_{xy}$ and the longitudinal 
conductivity $\sigma_{xx}$ versus energy $E$ for disorder strength $W/V=1$
({$\scriptstyle\Box$}, {\small +}) and $W/V=2$ ($\diamond$, $\times$). 
The inset shows the transition from QHE to insulator for $E/V=-0.8$ with the
crossing of $\sigma_{xx}$ ({\small +}) and $\sigma_{xy}$ ($\diamond$) at the 
criticaldisorder $W_c/V\approx 2.8$.}
\end{figure}

The small shift to lower energies of the first critical states has to be
compared to the descent of the critical states originating from the 
band center that are now visible a little above $E/V=-3$. 
Further increase of the disorder eventually brings the two minima 
together, thereby annihilating the last conducting state so that the system 
becomes an insulator.  
This scenario is in accordance with the results of the calculation of the 
thermodynamic localization length by Sheng and Weng \cite{SW97}.

The disappearance of current carrying states, 
starting at the highest Landau bands, can also be seen in Fig.~5. 
Here the influence of increasing disorder on the Hall and longitudinal 
conductivities is shown. The decay of the longitudinal conductivity near 
the critical energies coincides with an increase of $\sigma_{xx}$
in the plateau regions which in turn destroys the quantized Hall conductivity
that exhibits a monotonical decay with increasing disorder \cite{SW98}. 
This direct QHE to insulator transition is shown in the 
inset for energy $E/V=-0.8$. The crossing point 
$\sigma_{xx}^c=\sigma_{xy}^c\simeq 1.5$ lies at the critical disorder 
$W_c/V\approx 2.8$. For small $W$, both the conductivity $\sigma_{xx}$ and 
the resistivity $\rho_{xx}$ are zero while for $W>W_c$, $\sigma_{xx}\to 0$, 
but $\rho_{xx}$ tends to infinity. 

In conclusion, the QHE to insulator transition has been investigated in a
disordered tight binding model. From the energy level statistics the shift 
down in energy of states with negative Chern numbers becomes apparent. 
In contrast to the continuum model we find that direct transitions to the 
insulator are possible also from higher quantized Hall values. 
Our results support the scenario proposed recently by Sheng and Weng in 
Refs.~\cite{SW97,SW98}.


\end{document}